\documentclass[a4paper,11pt]{article}
\usepackage[all]{xy}
\usepackage{latexsym,amssymb,amsmath,amsfonts, amsthm, xcolor}

\usepackage[all]{xy}
\usepackage{latexsym,amssymb,amsmath,amsfonts, amsthm, xcolor}

\def\vbar{\mathchoice{\vrule height6.3ptdepth-.5ptwidth.8pt\kern-.8pt}
   {\vrule height6.3ptdepth-.5ptwidth.8pt\kern-.8pt}
   {\vrule height4.1ptdepth-.35ptwidth.6pt\kern-.6pt}
   {\vrule height3.1ptdepth-.25ptwidth.5pt\kern-.5pt}}
\def\fudge{\mathchoice{}{}{\mkern.5mu}{\mkern.8mu}}
\def\bbc#1#2{{\rm \mkern#2mu\vbar\mkern-#2mu#1}}
\def\bbb#1{{\rm I\mkern-3.5mu #1}}
\def\bba#1#2{{\rm #1\mkern-#2mu\fudge #1}}
\def\bb#1{{\count4=`#1 \advance\count4by-64 \ifcase\count4\or\bba A{11.5}\or
   \bbb B\or\bbc C{5}\or\bbb D\or\bbb E\or\bbb F \or\bbc G{5}\or\bbb H\or
   \bbb I\or\bbc J{3}\or\bbb K\or\bbb L \or\bbb M\or\bbb N\or\bbc O{5} \or
   \bbb P\or\bbc Q{5}\or\bbb R\or\bbc S{4.2}\or\bba T{10.5}\or\bbc U{5}\or
   \bba V{12}\or\bba W{16.5}\or\bba X{11}\or\bba Y{11.7}\or\bba Z{7.5}\fi}}

\newcommand{\vs}{\vspace{0.25cm}}

\newtheorem{theorem}{Theorem}
\newtheorem{itlemma}{Lemma}[section]
\newtheorem{itproposition}[itlemma]{Proposition}
\newtheorem{itcorollary}[itlemma]{Corollary}
\newtheorem{itremark}[itlemma]{Remark}
\newtheorem{itremarks}[itlemma]{Remarks}
\newtheorem{itdefinition}[itlemma]{Definition}
\newtheorem{itexample}[itlemma]{Example}

\newenvironment{lemma}{\begin{itlemma}\rm}{\end{itlemma}} %no-italics
\newenvironment{remark}{\begin{itremark}\rm}{\end{itremark}} %no-italics
\newenvironment{remarks}{\begin{itremarks} \rm}{\end{itremarks}}
\newenvironment{corollary}{\begin{itcorollary}\rm}{\end{itcorollary}}
\newenvironment{proposition}{\begin{itproposition}\rm}{\end{itproposition}}
\newenvironment{definition}{\begin{itdefinition}\rm}{\end{itdefinition}}
\newenvironment{example}{\begin{itexample}\rm}{\end{itexample}}
\newenvironment{fact}{\noindent {\em Fact}. \ \ }{\hfill \medskip}
\newenvironment{claim}{\noindent {\em Claim}. \ \ }{\hfill \medskip}
\newcommand{\be}[1]{\begin{equation}\label{#1}}
\newcommand{\ee}{\end{equation}}
\newcommand{\bl}[1]{\begin{lemma}\label{#1}}
\newcommand{\br}[1]{\begin{remark}\label{#1}}
\newcommand{\brs}[1]{\begin{remarks}\label{#1}}
\newcommand{\bt}[1]{\begin{theorem}\label{#1}}
\newcommand{\bd}[1]{\begin{definition}\label{#1}}
\newcommand{\bp}[1]{\begin{proposition}\label{#1}}
\newcommand{\bc}[1]{\begin{corollary}\label{#1}}
\newcommand{\bfact}[1]{\begin{fact}\label{#1}}
\newcommand{\bex}[1]{\begin{example}\label{#1}}
\newcommand{\ec}{\end{corollary}}
\newcommand{\efact}{\end{fact}}
\newcommand{\eex}{\end{example}}
\newcommand{\el}{\end{lemma}}
\newcommand{\er}{\end{remark}}
\newcommand{\ers}{\end{remarks}}
\newcommand{\et}{\end{theorem}}
\newcommand{\ed}{\end{definition}}
\newcommand{\ep}{\end{proposition}}
\newcommand{\epr}{\end{proof}}
\newcommand{\bpr}{\begin{proof}}
\newcommand{\bcl}{\begin{claim}}
\newcommand{\ecl}{\end{claim}}

\newcommand{\bi}{\begin{itemize}}
\newcommand{\ei}{\end{itemize}}
\newcommand{\ben}{\begin{enumerate}}
\newcommand{\een}{\end{enumerate}}
\begin{document}

\begin{center}

{\Large{Controllability of Quantum Walks on Graphs}}

%\author{Francesca Albertini\footnote{Dipartmento di Mathmatica Pura e Applicata, Universita'  di
%Padova, Italy\ \ Electronic adress:albertin@math.unipd.it}  and
%Domenico D'Alessandro\footnote{Department of Mathematics, Iowa State
%University, Ames, Iowa, U.S.A.\ \ Electronic address:
%daless@iastate.edu}}

\vs

%\maketitle

{\large{Francesca Albertini\footnote{Dipartmento di Mathmatica Pura
e Applicata, Universit\'{a }di  Padova, Via Trieste 63,  35121 Paodva, Italy;
Office Phone: +39-049-827-1376; Electronic
address: albertin@math.unipd.it}
 and Domenico D'Alessandro\footnote{Department of Mathematics, Iowa State
 University,  440 Carver Hall,  Ames IA-5001, Iowa, U.S.A.;
Office Phone: +1-515-294-8130;
Electronic address:
 daless@iastate.edu}}}

\end{center}

\vs
\begin{abstract}

In this paper, we consider discrete time quantum walks on graphs
with coin focusing on the  decentralized model, where the coin
operation is allowed to change with the vertex of the graph. When
the coin operations can  be modified at every time step, these
systems can be looked at as control systems and techniques of
geometric control theory can be applied. In particular, the set of
states that one can  achieve can be described  by studying
controllability. Extending previous results, we give a
characterization of the set of reachable states in terms of an
appropriate Lie algebra. We then prove general results and criteria
relating controllability to the algebraic and topological properties
of the walk. As a consequence of these results, we prove that if the
degree of the underlying graph is larger than $\frac{N}{2}$, where
$N$ is the number of nodes, the quantum walk is always completely
controllable, i.e., it is possible to having it to evolve according
to an arbitrary unitary evolution. Another result is that
controllability for decentralized models only depends on the graph
and not on the particular quantum walk defined on it. We also
provide explicit algorithms for control  and quantify the number of
steps needed for an arbitrary state transfer. The results of the
paper are of interest in
 quantum information theory where quantum walks are used and
 analyzed in the development of quantum algorithms.

\end{abstract}

%http://www.cs.bris.ac.uk/~montanar/presentations/QuantumWalks.ppt
%%%%%%%%%%%%%%%%%%%%%%%%%%%%%%%%%%%%%%%%%%%%%%%%%%%%%%%%%%%%%%%%%%%%%%%%%%%%%%%%
\vs


\vs

\section{INTRODUCTION}
In recent years, quantum walks on graphs have emerged as one of the
most useful  protocols to design quantum algorithms. This concerns,
in particular, problems that are naturally formulated on a graph,
such  as search problems where one is allowed to visit one location
at a time moving between neighboring vertices. The study of these
systems has now developed in a new rich area of quantum information
and mathematics. There are several aspects that are worth studying,
all interconnected: The design of quantum algorithms with better
performances than the classical ones, and in particular than the
randomized algorithms based on classical random walks; the
complexity theory of these algorithms; the dynamics of these
systems; their physical implementation. Reviews on quantum walks and
their algorithmic applications can be found in \cite{Ambainis},
\cite{Kempe}, \cite{Kendon}. Moreover, quantum walks are often used
as appropriate models in the study of natural phenomena (see, e.g.,
\cite{MITguy} for their application in a study of energy transfer in
photosynthesis).

\vs

There are two different versions of quantum walks, continuous and
discrete time. In its simplest form, a continuous time quantum walk
on a graph is a quantum system with state $\psi$ evolving according
to the Schr\"odinger equation \be{ConQW} i\dot \psi=H\psi,  \ee
where the Hamiltonian is constrained by the underlying graph, i.e.,
$h_{jk}\not=0$ if and only if there is an edge connecting the $j$-th
and $k$-th vertex of the graph. One important case is when $H$ is
the adjacency matrix of the graph. Discrete time quantum walks come
in different forms. One may use a quantum system, whose basis states
represent the edges of the graph and define the evolution on the
corresponding Hilbert space (see, e.g., \cite{MHillery} and
references therein) or one may use two quantum systems, called the
{\it coin} and the {\it walker}, the coin having dimension equal to
the degree $d$ of the graph (assumed regular) and the walker having
dimension equal to the number of vertices $N$. This second model,
although restricted to regular graphs, has the advantage of making
the role of the coin more transparent and intuitive and requiring a
Hilbert space whose dimension ($dN$) may be significantly smaller
than the one ($N^2$) for the walk defined on the edges of the graph.
There are some known relations among the various types of quantum
walks. Some of them  are discussed in \cite{Childs},
\cite{MikoConDis}.

\vs

In this paper,  we consider discrete time quantum walks with coin on
regular graphs. The evolution of these systems at every step is the
sequence of two operations; one operation on the coin system, called
{\it coin tossing},  and one operation on the walker system, called
{\it the conditional shift},  which changes the state of the walker
according to the state of the coin. We assume that, at every step,
one can change the coin tossing transformation and we adopt a
 {\it decentralized} model where the coin transformation may depend
 on the current state of the walker. This model is useful, for
 example, in search algorithms where there must be a way to
 distinguish one or more vertices from the other ones (see, e.g.,
 \cite{Ambaal} \cite{Tulsi}). The main topic  of this paper is to
 characterize the set of states that can be obtained with these
 models.

\vs

 The paper is organized as follows. In section \ref{MD} we
 describe in mathematical terms  the models that we want to
 study. In section \ref{C} we define the controllability of these
 models and give criteria to describe the set of reachable states.
 In particular, by  modifying the proof that was given in
 \cite{FD}, \cite{DGF} we extend and strengthen a result which
 describes the set of admissible evolutions of these systems as a Lie
 group. This Lie group might have one or more  connected components
 and its Lie algebra is generated by an appropriate set of matrices.
 An important problem, in this context,  is to characterize explicitly this
 Lie algebra for various quantum walks. A discussion is presented in section
 \ref{C} to further motivate this study. In section \ref{AlgTop},
 we relate the Lie algebraic controllability criterion described
 in section \ref{C} with the orbits of the permutations associated with the walk. This
 correspondence  will allow us to infer further properties of the
 controllability of these systems and in particular to solve the
 Lie algebra characterization problem above mentioned. As a
 consequence of general results we obtain several strong statements
 in special cases. In particular,
 quantum walks with graph of degree $d$ greater than $\frac{N}{2}$,
 with $N$ the number of vertices, are always completely
 controllable (this includes in particular complete graphs). Complete controllability
 means that every unitary evolution can be obtained with the
 dynamics of the system.
 We also identify the general
 structure for the Lie group of admissible evolutions. In section
 \ref{construct} we adopt a more direct approach to the study of
 controllability, by giving explicit constructive algorithms for
 state transfer. In doing this, we obtain an upper bound on the worst case number
 of steps needed for an arbitrary state transfer. In relating these
 results with the ones of the previous sections we notice that
 controllability only depends on the graph and not on the walk
 defined on it and that even purely graph theoretic questions can be
 answered using the concept of quantum walk (cf. Theorem \ref{independence}
 and the discussion that follows). Section \ref{deg2} contains
 some examples including a full
 treatment for graphs of degree two (i.e., cycles).
 % and a discussion on the impact
 %of the results obtained here on problems in quantum information.

\section{MODEL DEFINITION}
\label{MD}

Let  $G:=\{V, E \} $ be  a graph, where $V$ denotes the set of
vertices  of cardinality $N$ and $E$ the set of edges. We assume
that
\begin{itemize}
\item[H1)] $G$ is  a regular graph and we denote by $d$ its  degree.
\item[H2)] $G$ is connected and without self-loops.
\end{itemize}

We consider two quantum systems: a walker system whose state varies
in an $N$-dimensional space ${\cal W}$ (the {\it walker space}) and
a coin system whose state varies in a $d$ dimensional space ${\cal
C}$ (the {\it coin space}).  We denote by $|0 \rangle
,\ldots,|N-1\rangle,$ an orthonormal basis of the walker space
${\cal{W}}$ and   by $|c_1 \rangle,\ldots,|c_d\rangle$ an
orthonormal basis of the coin space ${\cal{C}}$. The meaning of the
state $|j\rangle$ is that if we measure the position of the walker
we find the position $j$ with certainty. Analogously, the meaning of
the state $|c_j \rangle$ for the coin is that  the ($d$-dimensional)
coin is giving the result $c_j$.

 With
this notation, we define a  {\it coin tossing operation} on ${\cal
C} \otimes {\cal W}$ as an operation of the type   \be{definition-1}
C:=\sum_{j=0}^{N-1} Q_j \otimes |j\rangle \langle j| ,\ee where
$Q_j\in U(d)$. This operation applies a unitary evolution to the
coin state which is allowed to depend on the current walker state.
This may be referred as a `decentralized' model as opposed to the
case where the coin evolution $Q_j$ does not depend on $j$, i.e., it
is the same for every walker state.  We also define a {\it
conditional shift} as an operator \be{definition-2} S:=\sum_{k=1}^d
|k \rangle \langle k| \otimes P_k,  \ee which applies to a state in
${\cal W}$ a permutation $P_k$ depending on the current value of the
coin system. In the basis $|c_k \rangle \otimes |j \rangle:=e_{kj}$,
$k=1,\ldots,d$, $j=0,\ldots,N-1$, $S$ has the matrix representation
\be{definition-2bis} S=\left(
\begin{array}{cccc}
    P_1  &  0   &  \cdots & 0 \\
    0   & P_2& \cdots &  0 \\
     0  & 0& \cdots & 0 \\
     \vdots & \vdots & \vdots & \vdots \\
     0 & 0 & \cdots & P_d \end{array} \right).
     \ee
The conditional shift $S$ has to be compatible with the graph
underlying the walk. This means that for every permutation $P_i$,
$i=1,\ldots,d$,  $P_i |j \rangle = |l\rangle$ implies that there
exists an edge in  $E$ connecting the vertices $j$ and $l$. Moreover
we will also have that for all $|j \rangle$, $i \not=k$ implies
$P_i|j \rangle \not= P_k |j \rangle$, which means that different
coin results have to induce different transitions on the graph. This
requirement also implies that, if there is an edge in $E$ connecting
$j$ and $l$ there must be a permutation $P_i$ such that $P_i |j
\rangle = |l\rangle$ and that the sum of the matrix representatives
of the permutations  $P_i$'s is the adjacency matrix of the graph.

Summarizing, the action of the coin tossing operation and
conditional shift on the vector space ${\cal C} \otimes {\cal W}$ is
given in  the basis $e_{ij}=|c_i>\otimes |j>$ $i=1,\ldots,d$, and
$j=0,\ldots,N-1$, by
\[
Ce_{ij}= (Q_j |c_i\rangle)\otimes |j\rangle,\]
\[
Se_{ij} = |c_i\rangle\otimes (P_i |j\rangle ). \]

The state of the quantum walk is described by a vector $|\psi
\rangle$ in ${\cal C} \otimes {\cal W}$, i.e.,
\[ |\psi \rangle :=
\sum_{k=1}^d  \sum_{j=0}^{N-1} \alpha_{kj} |c_k \rangle \otimes |j
\rangle.\]
 The probability of finding the walker in position $j$,
$p_j$ , is found by tracing out the coin degrees of freedom, that
is, $p_j=\sum_{k=1}^d |\alpha_{kj}|^2$.

\vs
The dynamics of the quantum walk is defined as
follows. At every step  $|\psi \rangle$ evolves as $|\psi \rangle
\rightarrow S C |\psi\rangle$, i.e., a coin tossing operation $C$ is
followed by a conditional shift $S$. The coin tossing operation may
change at any  time step preserving however the structure
(\ref{definition-1}). This leads to a point of view where the
operations $Q_j$ in (\ref{definition-1}) are seen as {\it control
variables} in the evolution of the system.

\section{CONTROLLABILITY}
\label{C} In this paper, we are interested in studying the set of
states that can be obtained for the quantum walks above defined by
varying in all possible ways the coin operations. The possible
evolutions are given by the set of all products of the form
$\prod_{k=1}^mSC_k$ where $C_k$ are arbitrary coin tossing
operations of the form (\ref{definition-1}). This set was already
studied in \cite{FD}, \cite{DGF} for the centralized case where the
$Q_j$ in (\ref{definition-1}) are all equal.
 Following the same
technique we obtain a characterization of this set in our case in
 Theorem \ref{ControllabilityTHEO}. We first set up some
definitions. Recall that $S$ being a permutation matrix has a
certain order $r$, such that $S^r$ is the identity on ${\cal C}
\otimes {\cal W}$. Define the set of matrices \be{calF} {\cal F}:=\{
{\cal A}, \, S {\cal A} S^{r-1}, \, \ldots,\,  S^{r-1} {\cal A} S\},
\ee where ${\cal A}$ is the set of matrices of the form
$\sum_{j=0}^{N-1} A_j \otimes |j \rangle \langle j|$ with $A_j \in
u(d)$. Notice that ${\cal A}$ is a Lie algebra, which is, in fact,
the direct sum of $N$ $u(d)$'s.\footnote{There are several
introductory books on Lie algebras and Lie groups (see e.g.,
\cite{Helgason}, \cite{SagleWalde}, \cite{Sepanski}). The book
\cite{Mikobook} presents introductory notions with a view to
applications to quantum systems.} Let ${\cal L}$ be the Lie algebra
generated by ${\cal F}$ defined as the smallest Lie algebra
containing ${\cal F}$ and let $e^{\cal L}$ be the connected Lie
group associated with ${\cal L}$, that is, the connected component
containing the identity. Consider the Lie group ${\bf G}$ generated
by $e^{\cal L}$ and $\{S \}$. This Lie group can be described in
different ways.

\bp{differentways}
Let:
\begin{enumerate}

\item ${\bf G}$ be the Lie group  generated by $e^{\cal L}$ and $\{ S
\}$.

\item  ${\bf K}$ be the set defined as:
\be{setB} {\bf K}:= e^{\cal L} \, \cup \,  e^{\cal L}S \, \cup \,
e^{\cal L}S^2 \, \cup \, \cdots \, \cup \,  e^{\cal L}S^{r-1} \ee
where $e^{\cal L}S^j$ is the set of all matrices $XS^j$ with $X \in
e^{\cal L}$.\footnote{Notice that this set is the same as the set of
all matrices $S^jY$ with $Y \in e^{\cal L}$. We can write $XS^j$ as
$S^jS^{r-j}XS^j$ and $S^{r-j}XS^j \in e^{\cal L}$ if $X \in e^{\cal
L}$ and the claim follows by defining $Y:=S^{r-j}XS^j$.}

\item If $p$ is  the smallest integer $1 \leq p \leq r$ such that $S^p \in e^{\cal L}$,
let   ${\bf C}$ be the set  defined as the disjoint union of
$e^{\cal L}$, $e^{\cal L}S$, ..., $e^{\cal L}S^{p-1}$.\footnote{To
see that this is a disjoint union, notice that  if there exists two
different indices $0 \leq k<j \leq p-1$ and two elements in $e^{\cal
L}$, $X$ and $Y$ such that $XS^j=YS^k$, we would have $S^{j-k} \in
e^{\cal L}$ which contradicts the minimality of $p$.}

\end{enumerate}

Then ${\bf G}={\bf K}={\bf C}$.
\ep
 \bpr It follows from the definitions that ${\bf K} \subseteq {\bf G}$, ${\bf
 C}
\subseteq {\bf G}$ and ${\bf C} \subseteq {\bf K}$. The claim
follows if we show that ${\bf G} \subseteq {\bf K}$ and ${\bf K}
\subseteq {\bf C}$. An element in ${\bf G}$ is a product
$\prod_{k=0}^m Y_k$, with $Y_0$ equal to the identity,  where $Y_k
\in e^{\cal L}$ or $Y_k=S$, for $k \geq 1$. By induction on $m$, if
$m=0$, this product is the identity which is in $e^{\cal L}$ and
therefore in ${\bf K}$. If $m>0$, write $\prod_{k=0}^m Y_k$ as
$Y\prod_{k=0}^{m-1} Y_k$, with $\prod_{k=0}^{m-1} Y_k \in {\bf K}$,
i.e., $\prod_{k=0}^{m-1} Y_k =XS^j$ for some $0 \leq j \leq r-1$ and
$X \in e^{\cal L}$. Now, if $Y \in e^{\cal L}$, then $YXS^j\in
e^{\cal L}S^{j} \subseteq {\bf K}$. If $Y=S$ then
$SXS^j=SXS^{r-1}S^{1}S^{j}$ and since $X \in e^{\cal L}$ implies
$Z:=SXS^{r-1} \in e^{\cal L}$, we have $YXS^{j}=ZS^{j+1} \in {\bf
K}$.

To see that  ${\bf K} \subseteq {\bf C}$, we need to consider only
$XS^k$ with $k
> p-1$. Choose $n$ so that $0\leq k-np \, \, \texttt{mod} \, r<p$.
 We have $XS^k=XS^{np}S^{k-np}:=YS^j$ with $Y=XS^{np}\in e^{\cal L}$
 and $j:= k-np \, \, \texttt{mod} \, {r}$ and this is in ${\bf C}$.
\epr

Notice that if $S \in e^{\cal L}$, ${\bf G}$ has only one connected
component which is given by $e^{\cal L}$. The following theorem
characterizes the controllability of the quantum walks.

\bt{ControllabilityTHEO} Let ${\cal E}$ be the set of possible
evolutions of the quantum walk. Then \be{theostatment} {\cal E}={\bf
G} \ee \et \bpr ${\cal E}$ is the set of products of transformations
of the form $SC$ with $C$ a coin tossing operation and $S$ a
conditional shift. Since $C \in e^{\cal L}\subseteq {\bf G}$ and $S
\in {\bf G}$ then $SC \in {\bf G}$ and therefore ${\cal E} \subseteq
{\bf G}$. Viceversa, consider the characterization of ${\bf G}$ as
${\bf K}$ in the above proposition and consider an element $XS^j \in
{\bf K}$, for some $0\leq j \leq r-1$. Since $X \in e^{\cal L}$ it
can be written as the product of matrices of the form $S^k e^{A}
S^{r-k}$ with $A$ a matrix of the form $A= \sum_{l=0}^{N-1} A_l
\otimes |l\rangle \langle l|$ and $A_l \in u(d)$. $e^{A}$ is a coin
operation $C$, and therefore, we can write $S^k e^{A} S^{r-k}$ as
$S^k C S^{r-k}$ and we can obtain it by performing $r-k$ steps with
coin operation equal to the identity, one step with coin operation
equal to $C$ and $k-1$ steps with coin operation equal to the
identity (in the case k=0, we can use one step with coin operation
equal to $C$ followed by $r-1$ operations with coin operation equal
to the identity). Therefore every matrix  of the form $S^k e^{A}
S^{r-k}$ can be obtained as an evolution of the quantum walk. So can
every product of such matrices and therefore every $X \in e^{\cal
L}$. To obtain $XS^j$, just compose the sequence giving $X$ with $j$
steps of the walk with coin operation equal to the identity. This
shows that ${\bf G} \subseteq {\cal E}$ and concludes the proof of
the theorem. \epr

An analogous characterization of the set ${\cal E}$ can be proved
with just small notational modifications for the `centralized' case
where all the matrices $Q_j$ in (\ref{definition-1}) are equal. In
this case, the Lie algebra ${\cal A}$ in (\ref{calF}) has to be
replaced by the Lie algebra of matrices $A \otimes {\bf 1}$ with $A
\in u(d)$ and ${\bf 1}$ the $N \times N$ identity. This was the case
treated in \cite{FD}, \cite{DGF}. The above discussion goes however
further with respect to the results in \cite{FD}, \cite{DGF} where
only the inclusion $e^{\cal L} \subseteq {\cal E}$ was proved.

{}From theorem \ref{ControllabilityTHEO}, it is clear that the Lie
algebra ${\cal L}$ plays a crucial role in the characterization of
the set of available state transformations  with the quantum walk.
Following common terminology in quantum control, we shall call this
Lie algebra the {\it dynamical Lie algebra} associated with the
quantum walk.  If ${\cal L}$ is $u(dN)$ the system is completely
controllable, that is every unitary operation can be obtained by
evolutions of the walk. We remark that this condition is also
necessary. If ${\bf G}=U(dN)$ then ${\bf G}$ can only have one
connected component since $U(dN)$ is  connected. Therefore $p=1$ in
Proposition \ref{differentways}.
 We can summarize this in the following theorem.
 \bt{Thecon}
The quantum walk is completely
controllable (every unitary operation
is possible) if and
only if ${\cal L}=u(dN)$.
\et

Another motivation to study the Lie algebra ${\cal L}$ is given by
the work in \cite{MikoConDis} where a procedure was given to obtain
the continuous quantum walk as an appropriate limit of a discrete
quantum walk. This procedure generalized a method given in
\cite{Strauch} for the quantum walk on the line. The set $i{\cal
L}$, represents the set of all Hamiltonians whose associated
continuous dynamics can be obtained with this procedure over the
full space ${\cal C} \otimes {\cal W}$. One then restrict oneself to
a smaller subspace to obtain a continuous quantum walk on a space
isomorphic to ${\cal W}$.

In the following section, we shall characterize the dynamical Lie
algebra ${\cal L}$ for every quantum walks in combinatorial terms,
i.e., in terms of the permutations $P_1$,...,$P_d$ characterizing
the walk.

\section{CONTROLLABILITY AND ORBITS OF PERMUTATIONS}
\label{AlgTop}

We now take a closer look at the generating set ${\cal F}$ in
(\ref{calF}) for the dynamical Lie algebra ${\cal L}$ and at how  it
relates to the orbits of the permutations $P_1$, ..., $P_d$ acting
on ${\cal W}$. We consider matrices in ${\cal F}$ (and ${\cal L}$)
in the basis $e_{ij}$ defined in section \ref{MD}. Consider a matrix
$S^kCS^{-k}$ in ${\cal F}$, for fixed $k$. We write
\be{SkCSmk}
S^{k}C S^{-k}= \left( \sum_{l=1}^d |l \rangle \langle l| \otimes
P_l^k \right)
  \left( \sum_{j=0}^{N-1} Q_j \otimes |j
\rangle \langle j| \right) \left( \sum_{m=1}^d |m \rangle \langle m|
\otimes P_m^{-k} \right)=\ee
\[
\sum_{\begin{matrix} l,m=1,\ldots, d  \\  j=0,\ldots,{N-1} \end{matrix}} | l \rangle
 \langle l |Q_j |m \rangle \langle m| \otimes P_l^k |j \rangle \langle
j|P_m^{-k}.
\]

After defining \be{defx} x_{jlm}:= \langle l | Q_j |m \rangle,  \ee
we can write
 \be{SKCfin} S^k C S^{-k}= \sum_{\begin{matrix}l,m=1,\ldots,d
\\ j=0,\ldots,N-1\end{matrix}} x_{jlm} |l \rangle \langle m | \otimes P_l^k |j
\rangle \langle j| P^{-k}_m.  \ee This expression tells us that, in
the $N \times N$ block determined by $l$ and $m$, the only non zero
terms are the ones corresponding to walker indices $r$ and $s$ such
that there exists a $j=0,1,\ldots,N-1$ with $r=P_l^kj$ and
$s=P_m^kj$. This means that the elements $(r,s)$ which are possibly
different from zero are such that $r=P_l^k P_m^{-k}s$, or,
equivalently, they are entries which are different from zero in the
permutation $P_l^k P_m^-k$.  Moreover from (\ref{defx}) these
entries are arbitrary complex numbers except for the requirement
that $S^kCS^{-k}$ is skew-Hermitian.\footnote{This can be easily
adapted if we are considering models different from the fully
decentralized one.} That is, $x_{jlm}$ in (\ref{defx}) are arbitrary
complex numbers except for the requirement that
$x_{jlm}^*=-x_{jml}$. Since $k$ is arbitrary, we obtain a
requirement for the entries of the matrices in ${\cal F}$ to be non
zero and arbitrary modulo the requirement that the matrix is
skew-Hermitian. This can be expressed in terms of the orbits of the
permutations $P_1,$ ...,$P_d$.

Given $l$ and $m$, $l,m=1,2,\ldots,d$ define the $l,m$-th joint
orbit ${\cal O}_{l,m}$ the subset of $V \times V$,
\be{jointorbit}
{\cal O}_{l,m}:= \bigcup_{\begin{matrix} k=0,1,\ldots,r-1 \\
j=0,1,\ldots,N-1 \end{matrix}} (P_l^kj,P_m^kj). \ee Notice that $(j,j)$ is in
any joint orbit for every pair $(l,m)$. In the basis given by
$e_{ij}$ we can enumerate the rows and columns of any matrix in
${\cal F}$ (and ${\cal L}$) using an index $i$ to identify a block
row (or column) ($i=1,2,\ldots,d$) and the index $j$
($j=0,1,\ldots,N-1$) to identify a position inside a block. This
discussion can be summarized as follows.

\bt{contorbit}
 The set ${\cal F}$ is the set of all the skew-Hermitian matrices
 having the $(l,r)-(m,s)$-th position $l,m=1,2,\ldots,d$, $r,s=0,1,\ldots,N-1$
 possibly different from zero if and
 only  if $(r,s) \in {\cal O}_{l,m}$.
\et

To study the nature of the Lie algebra generated by ${\cal F}$,
${\cal L}$, we shall now apply some results proved in
\cite{TuriniciRabitz}. We construct a {\it connectivity graph}
having $dN$ vertices each corresponding to a pair $(l,r)$, with
$l\in\{ 1,2,\ldots,d \}$ and $r \in \{0,1,\ldots,N-1\}$. We connect
two pairs $(l,r)$ and $(m,s)$ if and only if $(r,s) \in {\cal
O}_{l,m}$ that is, if and only if there is a matrix in ${\cal F}$
with the $(l,r),(m,s)$-th element different from zero. We omit the
self connections corresponding to diagonal elements, which can, in
fact, chosen arbitrarily (but must be purely imaginary). In
\cite{TuriniciRabitz} the authors studied the Lie algebra generated
by two skew-Hermitian matrices $H_0$ and $H_1$ with $H_0$ diagonal
and $H_1$, purely real, i.e., skew-symmetric, and with zeros on the
diagonal. A connectivity graph was associated with this pair with
edges connecting vertices corresponding to the row (or column)
indices $(a,b)$ if and only if the position $(a,b)$ in $H_1$ was
different from zero. These edges were then labeled, with the label
corresponding to $(a,b)$, $\omega_{ab}$ equal to
$|\lambda_a-\lambda_b|$, where $\lambda_a$ ($\lambda_b$) is the
diagonal element (eigenvalue) of $H_0$ corresponding to $a$ ($b$).
The result of \cite{TuriniciRabitz} we shall use is the following
\bt{Turrab} If the labeled connectivity graph is connected and it
remains connected after eliminating equal labels, then the system
\be{sistema} \frac{d}{dt} |\phi \rangle  =H_0 |\phi\rangle + H_1u
|\phi \rangle, \ee where $u$ is a control variable is state
controllable. \et State controllability in the previous statement
means that by varying the control $u$, it is possible to transfer
the state $|\phi \rangle $ between two arbitrary  values (with norm
equal to one). We now give a
 controllability condition based on the connectivity graph for the quantum walk.

\bt{connegraph} The quantum walk is completely controllable, i.e.,
${\cal L}=u(dN)$ if and only if the associated connectivity graph is
connected.
\et

\bpr First assume that the connectivity graph is connected. Since
${\cal F}$ contains arbitrary skew-Hermitian diagonal matrices we
can choose a matrix where all the differences between two diagonal
elements are different from each other. If we use this matrix with
the role of $H_0$ in Theorem \ref{Turrab} we obtain that the
associated differential system (\ref{sistema}) is state
controllable. This however does not necessarily imply that the
quantum walk is completely controllable, i.e., ${\cal L}=u(dN)$.
However, according to general controllability results for quantum
systems \cite{Notcontr} the only other possibility is that ${\cal
L}$ is conjugate to the symplectic Lie algebra $sp(\frac{dN}{2})$
plus multiples of the identity matrix. This implies that there
exists a matrix $\tilde J$ of the form $\tilde J:=T^\dagger J \bar
T$ where
\be{Jey} J=\begin{pmatrix} 0 & {\bf 1}_{\frac{dN}{2}} \\ {\bf
1}_{\frac{dN}{2}} & 0 \end{pmatrix},
\ee
and $T$ some unitary matrix, such that
\be{condsemisim} A \tilde J + \tilde J A^T=0,  \ee for every $A \in
{\cal L}$ with $Tr(A)=0$.\footnote{The fact that we have here $dN$
even is justified by the hand-shacking lemma of graph theory which
implies for regular graphs that $dN=2|E|$, where $|E|$ is the number
of edges.} However this is not possible. To see this, partition
$\tilde J$ into $d \times d$ blocks of dimension $N \times N$. {}From
the definition $\tilde J=T^\dagger J \bar T$, it follows that
$\tilde J^T=-\tilde J$. Formula (\ref{condsemisim}) has to hold for
every $A \in {\cal F}$, with $Tr(A)=0$ and in particular for any $d
\times d$-block skew Hermitian matrix with zero trace whose $N\times
N$ blocks are diagonal. Fix two block indices $k$ and $l$, in
$\{1,2,\ldots, d \}$. Taking all the blocks equal to zero except the
ones corresponding to the indices $k$ and $l$, equation
(\ref{condsemisim}) is equivalent to
 \be{equiv} \begin{pmatrix} D_{kk} &
D_{kl} \\ -D_{kl}^\dagger & D_{ll}\end{pmatrix}  \begin{pmatrix} \tilde J_{kk} & \tilde
J_{kl} \\ -\tilde J_{kl}^T & \tilde J_{ll}\end{pmatrix}+
  \begin{pmatrix} \tilde
J_{kk} & \tilde J_{kl} \\ -\tilde J_{kl}^T & \tilde
J_{ll}\end{pmatrix} \begin{pmatrix} D_{kk} & D_{kl} \\ -D_{kl}^\dagger & D_{ll}\end{pmatrix}=0,
\ee
 for any $D_{kk}$, $D_{kl}$, $D_{ll}$, $N \times N$ diagonal,
$D_{kk}$ and  $D_{ll}$ purely imaginary, and
$Tr(D_{kk})+Tr(D_{ll})=0$. Choosing $D_{kk}=i{\bf 1}$ and
$D_{ll}=-i{\bf 1}$ with $D_{kl}=0$, we obtain $\tilde J_{kk}=\tilde
J_{ll}=0$. Choosing $D_{kk}=D_{ll}=0$ and $D_{kl}={\bf 1}$ or
$D_{kl}=i{\bf 1}$, we obtain $\tilde J_{kl}^T=-\tilde J_{kl}$ or
$\tilde J_{kl}^T=\tilde J_{kl}$, respectively. Since $k$ and $l$ are
arbitrary, we obtain $\tilde J=0$ which is clearly not possible.
This shows that ${\cal L}=u(dN)$.

To see that the condition on the connectivity graph being connected
is also necessary, notice that if the graph is not connected then it
can be divided in $g \geq 2$ connected component. Reordering the
column and row indices of the matrices in ${\cal F}$, according to
the various connected components of the graph, we can write all the
matrices in ${\cal F}$ in block diagonal form. The Lie bracket
operation preserves this block diagonal form. Therefore, not all the
matrices in $u(dN)$ can be generated from the elements of ${\cal F}$
and ${\cal L} \not= u(dN)$.  \epr

Elaborating further on the statement and the proof of Theorem
\ref{connegraph} we obtain more information on the controllability
of quantum walks on graphs. In particular, notice that for every $j
\in V$, $(j,j)$ is in the orbit ${\cal O}_{lm}$ for every,
$l,m=1,2,\ldots,d$, which means that $(1,j)$, $(2,j)$, $\ldots$
$(d,j)$ are all connected in the connectivity graph. This means that
we can in fact consider a {\it reduced} connectivity graph whose
vertices correspond to the vertices of the original graph and there
is an edge connecting $r$ and $s$ if and only if there exist two
coin indices $l$ and $m$ so that $(l,r)$ and $(m,s)$ are connected
in the connectivity graph. In other terms, two vertices $r$ and $s$
in the reduced connectivity graph are connected by an edge if and
only if there exists a $j \in \{0,1,\ldots,N-1\}$ and two coin
indices $l$ and $m$ and an integer $k$ such that $P_l^kj=r$ and
$P_m^kj=s$, i.e., \be{redgraphconstruction} r=P_l^{-k}P_m^ks.   \ee
This relation gives a method to construct the reduced connectivity
graph. The algorithm is as follows

\vs

\vs

\vs

\vs

{\bf{Algorithm 1}}

\begin{enumerate}

\item Given the permutations $P_1$,...,$P_d$ characterizing the walk,
consider for every pair $l < m$ the permutations $P_l^{-k}P_m^k$
written in the cycle notation
$P_l^{-k}P_m^k=(\cdots)(\cdots)\cdots(\cdots)$.

\item Connect in a graph all the vertices that pairwise belong to the same cycle
at least in one instance. This is the reduced connectivity graph
associated with  the system.

%\item If, the system is
%completely controllable, that is, every unitary operation can be
%obtained to transfer from anyone state to any  other. Otherwise, any
%unitary operation can be obtained among states belonging to the same
%sets.

\end{enumerate}

\vs

In the case where the (unreduced) connectivity graph is not
connected, the connected components correspond to subsets of
vertices and when regrouping the row and column indices the
resulting matrices for every group still have the block form used in
(\ref{equiv}). Therefore the argument there can be repeated for
every single connected component and show that the Lie algebra
generated is the full unitary Lie algebra $u(dv)$ where $v$ is the
size (number of vertices) of the connected component in the reduced
connectivity graph. This shows that the general structure of the Lie
algebra ${\cal L}$ is as follows.

\bt{structureL} For a quantum walk, the dynamical Lie algebra ${\cal
L}$ is always the direct sum of $m >0$ Lie algebras\footnote{Direct
sum means that all these Lie algebras are summed in the vector space
sense and they all commute with each other.}  isomorphic to
$su(dv_j)$ for some positive integers $v_j$, $j=1,\ldots,m$ with
$\sum_{j=1}^m v_j=N$ and a one dimensional Lie algebra spanned by
multiples of the identity matrix in $u(dN)$. Each subalgebra
isomorphic to $su(dv_j)$ corresponds to a connected component of the
reduced connectivity graph with $v_j$ vertices. Complete
controllability is obtained in the case $m=1$. \et

\vs

In the rest of this section  we give  two consequences of the
results and methods summarized in Theorems \ref{connegraph} and
\ref{structureL} and Algorithm 1. Appendix A contains some further
analysis which is uses the results of the next section to show that
the number $m$ in Theorem (\ref{structureL} can only be 1
(controllable case) or 2.

\vs

%\bc{coro1} For quantum walks on graphs complete controllability and
%state controllability are equivalent. \ec \bpr If the system is not
%completely controllable then ${\cal L}$ has the structure of the
%previous proposition and this implies that it cannot be isomorphic
%to $sp(\frac{dN}{2})$ (plus multiples of the identity)
%\cite{Notcontr}. This is due to the fact that  $sp(\frac{dN}{2})$ is
%a simple Lie algebra and so has no nontrivial ideal while the Lie
%algebras isomorphic $su(dv_j)$ described in proposition
%\ref{structureL} are all ideals of ${\cal L}$. \epr
% More concretely, in the non controllable case, the Hilbert space
% ${\cal C} \otimes {\cal W}$ splits into $m$ components of the form
% ${\cal C} \otimes {\cal W}_j$, with $\bigoplus_{j=1}^m {\cal
% W}_j={\cal W}$ which are all invariant under the effect of the
% allowed dynamics of the walk. Therefore state controllability is
% not possible.

\vs

 \bp{corodN2} If $d >\frac{N}{2}$ the quantum walk is completely
 controllable.
\ep

\bpr As we have seen above, in the connectivity graph, elements
$(l,j)$ where $l$ is the coin index and $j$ the walker index, for
fixed $j$  are in the same connected component. For this reason we
considered a reduced controllability graph. Consider now the walker
index $1$. We have that $(1,P_1 1)$, $(2,P_2 1)$,\ldots,$(d,P_d1)$
are all connected in the connectivity graph. This means that in the
reduced connectivity graph vertices $P_11$, $P_21$,\ldots, $P_d1$
are all connected. {}From the condition $P_i1 \not=P_l1$ if $i \not=l$
we get that there are $d$ different vertices connected  in the
reduced connectivity graph. Consider now the walker index $2$. We
have that $(1,P_1 2)$, $(2,P_2 2)$,\ldots,$(d,P_d2)$ are all
connected in the connectivity graph. Therefore, in the reduced graph
$P_12$, $P_22$,\ldots, $P_d2$, which are all different,  are all
connected. Since $d > \frac{N}{2}$, the sets
$\{P_11,P_21,\ldots,P_d1\}$ and $\{P_12,P_22,\ldots,P_d2\}$ must
have an element in common. Therefore, the corresponding vertices in
the reduced connectivity graph are all connected. Proceeding this
way, we find that all vertices $P_lj$, for $l=1,\ldots,d$ and
$j=0,1,\ldots, N-1$ are connected in the reduced connectivity graph.
Since for every $ k\in \{0,1,\ldots,N-1\} $ there exist $j  \in
\{0,1,\ldots,N-1\} $ and $l \in \{1,2,\ldots,d\} $ such that
$k=P_lj$, the (reduced) connectivity graph is connected and the
quantum walk is completely controllable.

\epr

The bound in Proposition \ref{corodN2} is sharp in the sense that
there are quantum walks that are not controllable with
$d=\frac{N}{2}$. In fact, we shall see in section \ref{deg2} that
quantum walks on a cycle (therefore of degree $2$) with $4$ vertices
are not controllable. Notice also that, as a special case of
Proposition  \ref{corodN2}, quantum walks on complete graphs are
always controllable.\footnote{We always assume $N>2$.}

\vs

For the last result of this section, we need the concept of {\it
product} of two quantum walks. Consider two quantum walks the first
one, $W_1$  supported by a graph $G_1:=\{V_1, E_1\}$ with a set of
permutations $\{P_1,\ldots,P_{d_1}\}$ and the second one $W_2$
supported by a graph $G_2:=\{V_2, E_2\}$ with a set of permutations
$\{Q_1,\ldots,Q_{d_2}\}$. The product walk $W_1 \times W_2$ is the
walk whose graph is the Cartesian product of $G_1$ and $G_2$ and the
associated permutations are $\{\tilde P_1,\ldots, \tilde
P_{d_1},\tilde Q_1,\ldots, \tilde Q_{d_2} \} $ acting on the
vertices $(j,k) \in V_1 \times V_2$ as $\tilde P_l (j,k):=(P_l
j,k)$, $\tilde Q_l (j,k):=( j, Q_l k)$. One example is a walk on a
 $2$-dimensional lattice with $N_1 \times N_2$ vertices connected in a periodic
fashion horizontally and vertically. Coin results can be labeled
$R$, $L$, $U$, $D$ (Right, Left, (mod $N_1$), Up, Down (mod $N_2$),
respectively) and this is the product of two cycles one evolving
 horizontally on a cycle with $N_1$ nodes and one evolving
 vertically on a cycle with $N_2$ nodes.

\bp{propoprod}  The product of two controllable walks is
controllable. \ep

\bpr With the above notations, since the walk $W_1$ is controllable,
for every $j \in V_2$ the vertices $(k,j)$, $k=1,\ldots, N_1$ are
all connected in the reduced connectivity  graph. Analogously, from
the controllability of $W_2$, it follows  that for every $k \in V_1$
the vertices $(k,j)$, $j=1,\ldots, N_2$ are all connected in the
reduced connectivity graph. Therefore this graph is connected.
\epr

We remark that the above condition is not necessary and one can find
two quantum walks with one or both of them uncontrollable whose
product is controllable.

%Choose cycle with 4 vertices and product it with cycle with three vertices
%or two cycles with four vertices.
%

\section{CONSTRUCTIVE CONTROLLABILITY ALGORITHMS}
\label{construct}

In this section, we discuss the constructive controllability. We
will focus on finding control algorithms to steer the state of the
quantum walk between two values. Thus, for any given two state
vectors $|\psi_1 \rangle, |\psi_2 \rangle$ in ${\cal C} \otimes
{\cal W}$ we will {\em{find}} a
 sequence of coin tossing operations $C_1, \ldots,
C_k$, such that
\[
|\psi_2 \rangle = SC_k\cdots SC_1 |\psi_1 \rangle.
\]
Moreover, we will give a {\em{bound}} on the length $k$ of the
needed control sequence. Whether such a sequence exists or not can
be checked with the methods of the previous two sections.
%In
%particular notice that,  according to Corollary \ref{coro1}, state
%controllability and complete controllability are equivalent
%properties.

First,  we define, for a given node $j$, the set of all nodes that
one can  reach using the edges of the graph in a given number of
steps. Fix a node $j\in \{0,\ldots, N-1\}$, let:
\be{nodi-raggiungibili}
\begin{array}{ccl}
{\cal N}^0(j)  & := & \{ j \}, \\
{\cal N}^{k+1}(j) & := & \{ P_s(l) \, |\, l\in {\cal N}^k(j), \
1\leq s\leq d \}.
\end{array}
\ee With these definitions, $l \in {\cal N}^k(j)$ means that there
exists a sequence of permutations $R_1,\ldots,R_k$ in the set
$\{P_1,\ldots,P_d\}$ such that $l =R_k R_{k-1}\cdots R_1 j$. The
connectedness  assumption on the graph $G$ implies that $\forall \,
i,\, j\in \{0,\ldots, N-1\}$ there exists a $ k\geq 0$ such that
$i\in  {\cal N}^{k}(j) $. The set ${\cal N}^k(j)$ only depends on
the graph. It is the set of vertices which are connected to $j$ by a
path of length $k$.

{}From these observations, we can collect two properties of the sets
${\cal N}^{k}(j)$ in the next lemma.

\bl{lemma-nodi} Let $i,\, j, \, l \in \{0,\ldots,N-1\}$, $k,\, s\geq
0$,  we have:
\begin{enumerate}
\item
$l\in {\cal N}^{k}(j) \Leftrightarrow j \in {\cal N}^{k}(l),$
\item
if $l\in {\cal N}^{k}(j)$ and $i \in {\cal N}^{s}(j)$ then $i\in
{\cal N}^{k+s}(l)$.

\end{enumerate}
\el

\vs

Choose a node $j\in  \{0,\ldots, N-1\}$ and consider a state
$|\psi_1\rangle$ with probability $1$ to find the walker in this
position. Thus $|\psi_1 \rangle$ is of the form
  $|\psi_1 \rangle= |c \rangle\otimes |j\rangle$, for some state
$|c\rangle \in {\cal C}$. If there exists a sequence of coin tossing
operations of length $k$ such that
\[
SC_k  \cdots SC_1 |\psi_1 \rangle= \sum_s\sum_l
\alpha_{ls}|c_{k_l}\rangle\otimes |j_{k_s}\rangle.
\]
then,   $j_{k_s}\in {\cal N}^{k}(j)$ for all $k_s$. This fact, in
particular, implies that a necessary condition to have complete
controllability is that $\forall \, j\in \{0, \ldots, N-1\}$  there
exists a $k \geq 0$ such that  ${\cal N}^{k}(j)= \{0, \ldots, N-1\}
$ since we have to be able to transfer to an {\it arbitrary} state
in ${\cal C} \otimes {\cal W}$. By using property 2) of Lemma
\ref{lemma-nodi},
 we can substitute $\forall$ with $\exists$ in the previous
 sentence. In fact, if there exists a $\bar j$ such that with a path
 of length $k$, we can reach any $l \in \{0,1,\ldots, N-1 \}$, with
 a path of length $2k$ we can go from any $j \in \{0,1,\ldots, N-1
 \}$ to any $l \in \{0,1,\ldots, N-1 \}$ (just go to $\bar j$ in $k$
 steps and then to $l$ in $k$ additional  steps).

 Thus, we get that:

\bcl C1
\begin{quote}
 complete controllability $\Rightarrow$
 $\exists \, j\in \{0, \ldots, N-1\}$ and  $k\geq 0$
such that \\  ${\cal N}^{k}(j)= \{0, \ldots, N-1\} $.
\end{quote}
\ecl

This necessary condition can be checked indirectly with the methods
of the previous sections. The constructive algorithms we are going
to describe will imply that this necessary condition is indeed
sufficient to get controllability between two arbitrary states for
our models. Moreover our results will imply   an upper  bound on the
number of steps needed for arbitrary state transfer in terms of the
maximal (over $j$) ${k}$ such ${\cal N}^{{k}}(j)= \{0, \ldots, N-1\}
$ and of the order $r$ of the conditional shift matrix $S$.

The next proposition provides a first $k$-steps control algorithm to
go from a state with probability $1$ in a given  node $j$, i.e., a
state of the type  $ |c_0\rangle\otimes |j\rangle$, to one where the
probability is arbitrarily distributed on the nodes in ${\cal
N}^k(j)$.  Even if the proof of the next proposition, as well as the
proof of Proposition \ref{datuttiauno}, will be given by induction,
 they are  constructive. We present an  example in Section \ref{esempio-algoritmo}.

\bp{prima-nodi} Let $j$ be any node and  $A_k=\{v_1,\ldots,v_l\}$ be any subset of
 ${\cal
N}^{k}(j)$.  Fix  any state of the type
$|\psi_0\rangle= |c_0\rangle\otimes |j\rangle$ and any complex
coefficients $(\alpha_1, \ldots, \alpha_l)$ with $\sum_{a=1}^l
|\alpha_a|^2=1$ on the nodes of $A_k$. Then it is always
possible to construct a control sequence $C_1,\ldots, C_k$ of coin
operations such that: \be{coeffic} SC_k\cdots SC_1 |\psi_0 \rangle=
\sum_{h=1}^{l} \alpha_h |c_{h}\rangle\otimes |v_h \rangle, \ee for
some values of the coin variables $c_{h}$ (not necessarily
distinct). \ep
 \bpr
  We will prove
the statement by induction on $k$.

If $k=0$, then the statement is
obvious. Assume that the proposition holds for $k$.

 Let $A_{k+1}=\{v_1,\ldots,v_l\} \subseteq {\cal
N}^{k+1}(j)$. By definition of ${\cal N}^{k+1}(j)$ we have that
$A_{k+1}=\{v_1,\ldots,v_l\}  =\{R_{1}(w_{1}),R_{l}(w_{l})\}$,
where for $i=1,\ldots,l$  $w_{i}\in {\cal N}^{k}(j)$, and
$R_1,\ldots R_l$ are  permutations in the set $\{ P_1,\ldots,P_d
\}$. The nodes $w_h$ need not to be different. Denote by $s$ the
cardinality of $\{w_1,\ldots,w_l\}$, and let $A_k=
\{w_1,\ldots,w_l\} =\{z_1,\ldots,z_s\}$
 where all elements are distinct in the second set notation. Without
 loss of generality, we assume that we have ordered the nodes
$v_h \in A_{k+1}$ in such a way that the first $g_1$ of
$w_i$ are equal to $z_1$, the second $g_2$ of $w_i$ are equal to
$z_2$ and so on; so we have:
\[
\begin{array}{c}
z_1=w_{1}=\cdots=w_{{g_1}}, \\
z_2=w_{g_1+1}=\cdots=w_{g_1+g_2}, \\
\vdots\\
z_s=w_{g_1+\cdots+g_{s-1}+1}=\cdots=w_{g_1+\cdots+g_s},  \end{array}
\]
with $g_0:=0$.  Moreover denote by $c_h$ the coin value that
correspond to the transition from $w_h$ in ${\cal N}^k(j)$ to $v_h$
in ${\cal N}^{k+1}(j)$, i.e., \be{impo4} P_{c_h}w_h=v_h. \ee
 Let $\alpha_1,\ldots,\alpha_l$ be the
given coefficients (cf. (\ref{coeffic})) \footnote{We can assume
these coefficients  all different from zero, without loss of
generality,  as in the case where one of them is zero we can just
eliminate the corresponding $v_h$ from the sum (\ref{coeffic}).},
satisfying  $\sum_{h=1}^l |\alpha_h|^2=1$.

Define for $i=1,\ldots,s$,
\be{gammai}
\gamma_i:=\sqrt{\sum_{h=g_1+\cdots+g_{i-1}+1}^{g_1+\cdots+g_i}
|\alpha_h|^2}.
\ee
By the inductive assumption, since $A_k$ is a subset of ${\cal N}^{k}(j)$, it is
possible from $|\psi_0\rangle= |c_0\rangle\otimes |j\rangle$,  to
construct a sequence of $k$  coin operations that steers $|\psi_0
\rangle$   to:
\[
|\tilde{\psi}\rangle= \sum_{i=1}^{s}\gamma_i
|\delta_i\rangle\otimes|z_i\rangle,
\]
for some states  of the coin $|\delta_i\rangle$. Let $Q_{z_i}$ be
any unitary matrix such that:

\be{operazioni}
Q_{z_i} |\delta_i\rangle:= \frac{1}{\gamma_i}
\sum_{h=g_1+\cdots+g_{i-1}+1}^{g_1+\cdots+g_i} \alpha_h
|c_{h}\rangle,
\ee
where the $|c_{h}\rangle $ are the ones defined in (\ref{coeffic})
and the $\gamma_i$'s  are all different from zero because so are the
$\alpha_h$'s.

Define a coin tossing operation ${C}_{k+1}$ as the matrix where for
the nodes $z_i$ we use the previous matrix $Q_{z_i}$, and for the
other we use an arbitrary $Q$ in $U(d)$, e.g., the identity. We
have:

\[
S {C}_{k+1}(|\tilde{\psi}\rangle)=S\left( \sum_i \gamma_i (Q_{z_i}
|\delta_i\rangle)\otimes |z_i\rangle\right)=
S\left(\sum_i\left(\sum_{h=g_1+\cdots+g_{i-1}+1}^{g_1+\cdots+g_i}
\alpha_h |c_{h}\rangle \right) \otimes |z_i\rangle\right) =
\]
\[
=
 S\left(\sum_{h=1}^l \alpha_h |c_{h}\rangle\otimes |w_{h}\rangle
\right)=
 \sum_{h=1}^l \alpha_h |c_{h}\rangle\otimes
|P_{i_h}(w_{h})\rangle=
 \sum_{h=1}^l \alpha_h |c_{h}\rangle\otimes |v_{h}\rangle,
\]
as desired. In the last equality, we used (\ref{impo4}). \epr

The next  proposition shows how to reach a state of the form in the
right hand side of (\ref{coeffic}) where the $|c_h\rangle$ are
replaced by an arbitrary superposition of coin states.

\bp{daunoatutti} Let $j$ be any node, assume that ${\cal
N}^{k}(j)=\{v_1,\ldots,v_l\}$, and fix any state of the type
$|\psi_0\rangle= |c_0\rangle\otimes |j\rangle$. Then in at most
$k+r$ steps (where $r$ is the order of the conditional shift matrix
$S$),  we can reach, from $|\psi_0\rangle$, any state of the type
$|\psi_f\rangle = \sum_{h=1}^{l} \sum_{s=1}^d \alpha_{hs}
|c_s\rangle\otimes |v_h\rangle$ for arbitrary coefficients
$\alpha_{hs}$ such that $\sum_{h=1}^{l} \sum_{s=1}^d
|\alpha_{hs}|^2=1$.  \ep
\bpr Define $\beta_h=\sum_{s=1}^d
|\alpha_{hs}|^2$. We can assume, without loss of generality, that $\beta_h\neq 0$.
In fact, if $\beta_h=0$, then necessarily $\alpha_{hs}=0$ for all $s=1,\ldots,d$, and so in this case we can just eliminate $|v_h\rangle$ from the sum that defines $|\psi_f\rangle$.  {}From Proposition \ref{prima-nodi} we have a
sequence of $k$ coin operations $C_1,\ldots,C_k$ such that:
\[
SC_k\cdots SC_1 |\psi_0 \rangle= \sum_{h=1}^{l} \beta_h
|c_{h}\rangle\otimes |v_h\rangle,
\]
for some values of the coin variables $c_{h}$. Let $Q_{v_h}$ be any
unitary matrix such that:
\[
Q_{v_h} |c_{h}\rangle := \frac{1}{\beta_h}\sum_{s=1}^d \alpha_{hs}
|c_s \rangle
\]
Choose a coin tossing operation ${C}_{k+1}$ as the matrix where in
the nodes $v_h$ we use the previous matrix $Q_{v_h}$, and in the
other nodes we use an arbitrary $Q$ in $U(d)$. Letting
$C_{k+2}=\cdots=C_{k+r}=I$,
 we have:
\[
SC_{k+r}\cdots SC_1 |\psi_0 \rangle= S^r C_{k+1}\left(
\sum_{h=1}^{l} \beta_h |c_{h}\rangle\otimes |v_h\rangle\right) =
\]
\[ \sum_{h=1}^l
\beta_h (Q_{v_h} |c_{i_h}\rangle)\otimes |v_h\rangle=
\sum_{h=1}^{l} \sum_{s=1}^d
\alpha_{hs} |c_s\rangle\otimes |v_h\rangle,
\]
as desired.
\epr

\br{minimizer} In some cases one can choose values $\tilde C_1$ and
$\tilde C_2$ for the coin transformations so that \be{lpas} \tilde
C_2 S \tilde C_1=S^{-1}. \ee In these cases, we can replace
$C_{k+1}$ above with $\tilde C_1 C_{k+1}$ and $C_{k+2}=I$ with
$\tilde C_2$ and omit all the following steps to have
$SC_{k+2}S\tilde C_{1}={I}$ in the proof of the above theorem. In
these cases, one can replace $r$ with $2$ in  the statement of the
above theorem. \er

\vs

The previous propositions have shown how to go from a state with
walker state in a single node $j$ to a state where the  walker is
distributed according to an arbitrary superposition of states $v \in
{\cal N}^k(j)$. The following proposition shows how to perform the
converse type of state transfer.

\bp{datuttiauno} Let $j$ be any node, let  ${\cal
N}^{k}(j)=\{v_1,\ldots,v_l\}$, and fix  any state of the form

\be{formstate}
 |\psi_0\rangle = \sum_{h=1}^{l} \sum_{s=1}^d
\alpha_{hs} |c_s\rangle\otimes |v_h\rangle \ee
 for arbitrary
coefficients $\alpha_{hs}$ such that $\sum_{h=1}^{l} \sum_{s=1}^d
|\alpha_{hs}|^2=1$.  Then there exists a sequence of coin operations
of length   at most $k$ that steers the initial state
$|\psi_0\rangle$ to a state of the type
$|\psi_f\rangle=\sum_{s=1}^{d}\gamma_s |c_s\rangle \otimes
|j\rangle$. \ep \bpr As in proposition \ref{prima-nodi},  we will
prove the statement by induction on $k$.

If $k=0$, then the statement is
obvious. Assume that the proposition holds for $k$.

 Let ${\cal
N}^{k+1}(j)=\{v_1,\ldots,v_l\}=\{P_{1}(w_{1}),\ldots,P_{l}(w_{l})\}$,
where $w_{h}\in {\cal N}^{k}(j)$. Notice that, for all
$h=1,\ldots,l$, since $P_{h}(w_h)=v_h$, there exists also a coin
value $c_{j(h)}$ such that $P_{j(h)}(v_h)=w_h$. Let:
\[
\gamma_h:=\sqrt{\sum_{s=1}^d|\alpha_{hs}|^2},
\]
where $\alpha_{hs}$ are the ones defined in (\ref{formstate}).
We can assume $\gamma_h\neq 0$, otherwise we can just eliminate
$| v_h \rangle$ from the sum in equation (\ref{formstate}).  Let
$C_1$ be a coin tossing operation \be{C1cointoss} C_1:=\sum_{h=1}^l
Q_{v_h} \otimes |v_h \rangle \langle v_h| + Q \otimes (I-
\sum_{h=1}^l |v_h \rangle \langle v_h|),  \ee where $Q$ is any
arbitrary unitary on the coin space ${\cal C}$ and
\[
Q_{v_h}\left( \frac{1}{\gamma_h}\sum_{s=1}^{d} \alpha_{hs}
|c_{s}\rangle \right)= |c_{j(h)}\rangle.
\]
Then we have:
\[
SC_1 \left( \sum_{h=1}^{l} \sum_{s=1}^d \alpha_{hs} |c_s
\rangle\otimes |v_h\rangle \right)= S \left(  \sum_{h=1}^{l}
\gamma_h
 |c_{j(h)} \rangle\otimes |v_h\rangle \right) =
   \sum_{h=1}^{l} \gamma_h
 |c_{j(h)} \rangle\otimes |w_h\rangle.
\]
This concludes  the inductive step, since the nodes $w_1,\ldots,
w_h$ are in ${\cal N}^{k}(j)$.
\epr

The previous results show that it is possible to go from a state of
the form $|\psi_0\rangle :=|c_0\rangle \otimes |j\rangle$ to any
state of the form (\ref{formstate}) where the  $v_h$'s are in ${\cal
N}^k(j)$ and viceversa. If there exists a $j$ such that ${\cal
N}^k(j)=\{0,1,\ldots,N-1\}$, then the state in (\ref{coeffic})-(\ref{formstate})
 is just an arbitrary state and we can go from an
arbitrary state to a state of the form $|\psi_0\rangle = |c_0
\rangle \otimes |j\rangle $ in $k$ steps and from this state to an
arbitrary state in $k+r$ steps. Therefore every state transfer is
possible and it takes at most $2k+r$ steps. This gives the promised
converse of the Claim C1 and gives an upper  bound on the number of
steps needed for an arbitrary state transfer. This bound can be
sharpened by noticing that if there is a $j$ such that ${\cal
N}^k(j)=\{0,1,\ldots,N-1\}$ for some $k$ then this is true for every
$j$ with a $k$ which will in general depend on $j$. Therefore we
denote by $k_j$ the smallest $k$ such that ${\cal N}^k(j)=\{
0,1,\ldots,N-1 \}$. We can sharpen the previous upper bound on the
number of steps by choosing the $j$ such that $k_j$ is minimum. In
particular, define \be{defkappa} {\bf{k}}:= \min\left\{ \, k_j \, |
j \, \in V \, \right\}. \ee

We summarize the previous discussion in the following Theorem.

\bt{con-constructive} If a quantum walk is completely controllable
then there exists  a node $j$ such that ${\cal
N}^{k_j}=\{0,1,\ldots,N-1\}$, for some finite $k_j$. In that case
the property is true for every $j$. Viceversa if such a $j$ exists,
we can transfer between two arbitrary states ({\it state
controllability}). In this case, define ${\bf k}$ as in
(\ref{defkappa}). Let $r$ be the order of the conditional shift
matrix $S$. Then any state transfer can be performed in at most $2
{\bf k} +r$ steps.
\et

\vs

The previous theorem \ref{con-constructive} presents a gap between
two notions of controllability complete controllability and state
controllability which are in general not equivalent \cite{Notcontr}.
In order to fill this gap and have a perfect if and only if
condition in our theorem, we need to stud more closely  the relation
between the condition on ${\cal N}^k(j)$ and the condition of
Theorem \ref{structureL}. In doing this we will get more information
on the controllability of quantum walks and it will follow that the
two notions are in fact equivalent for the models we are
considering.

In Theorem \ref{structureL}, we partitioned the set of vertices of
the graph $G$ into subsets and divided the dynamical Lie algebra
${\cal L}$ into a certain number of subalgebras each one
corresponding to one of these subsets. In particular if there is
only one set the Lie algebra is the full Lie algebra $u(dN)$ and the
system is completely controllable. We now notice that two vertices
$w$ and $s$ are in the same subset (i.e., in the same connected
component of the reduced controllability graph if and only if there
exists a sequence of permutations of the form $P_l^kP_m^{-k}$, with
$l,m\in \{1,2,\ldots,d\}$ and some $k=0,1,2,...$ transferring $w$ to
$s$. This is equivalent to the fact that there exists a sequence of
permutations {\it of even length} transferring $s$ to $w$. To see
this first assume that
\begin{equation}\label{ppsd}
w=\prod_j P_{l_j}^{k_j}P_{m_j}^{-k_j}s.
\end{equation}
For any $y\in V$ and any $P_m$ $y$ and $P_m^{-1}y$ are connected in
the graph $G$. This means that there exists a $P_l$ such that
$P_m^{-1}y=P_ly$. Therefore we can replace every permutation with a
negative power with a (possibly different) permutation with positive
power in (\ref{ppsd}) and obtain our claim. Viceversa if
\begin{equation}\label{ppsd2}
w=\prod_t P_{l_t}P_{m_t}s,
\end{equation}
we can replace all the permutations $P_{m_j}$ with negative powers
of permutations and obtain an expression of the form (\ref{ppsd}).
Notice that this also shows that we can restrict ourselves to
considering $k_j=1$ in using (\ref{ppsd}) and partitioning the set
$V$. In view of these considerations complete controllability is
verified if and only for  any two nodes $w$ and $s$ there exists a
sequence of permutations of even length mapping $s$ in $w$. Now
assume that this is the case and fix a $j \in V$. Then for any $w
\in V$ there exists a sequence of even length mapping $j$ to $w$.
Let $2k_w$ this length depending on $w$ and let $2k_{max}$ the
maximum length, maximized over the $w$'s. We can go from $j$ to any
$w\in V$ in {\it exactly} $2 k_{max}$ steps, we just follow the path
with the given permutations for $2 k_w$ steps and then `oscillate'
back and forth with any neighbor $k_{max}-k_w$ times. Therefore
controllability implies that given $j$, there exists a $k=k(j)$
(even) such that we can reach any vertex in $V$ in exactly $k(j)$
steps on the graph (i.e., with a sequence of permutations of length
$k(j)$). Viceversa, if given $j$ there exists a $k(j)$ such that for
any $w$ there exists a sequence of length $k$ $P_{l_1}\cdots
P_{l_k}$ mapping $j$ to $w$ we have for $w$ and $s$, from
$w=P_{l_1}\cdots P_{l_k}j$, $s=P_{m_1}\cdots P_{m_k}j$
\begin{equation}\label{p2p4}
w=P_{l_1}\cdots P_{l_k}P_{m_k}^{-1} \cdots P_{m_1}^{-1}s,
\end{equation}
using the above argument to replace negative powers with positive
ones, that we can map any $s$ to any $w$ with a sequence of even
length of permutations and the system is completely controllable.
This shows the following.

\bt{equivale} The condition  ${\cal N}^k(j)=\{0,1,\ldots,N-1\}=V$ of
Theorem \ref{con-constructive} and the connectivity of the reduced
connectivity graph of Theorem \ref{structureL} are equivalent
properties. In particular state controllability and complete
controllability are equivalent notions for discrete time quantum
walks.
 \et

An important consequence
 of the controllability  criterion given in this
 section is that although the quantum walk and the concept
 of controllability where studied in connection with the defining
 permutations $\{P_1,\ldots,P_d\}$, we have the following.

\bt{independence} Controllability of a quantum walk on a graph only
depends on
 the topology of the graph and not on the particular permutations
 $\{P_1,\ldots,P_d\}$.
\et

In view of this result and the equivalence of the controllability
criteria given in this section and in the previous sections stated
in theorem \ref{equivale}, one may neglect the concept of
controllability and use the criterion of Theorem \ref{structureL} to
carry over graph theoretic analysis. In particular, given a regular
graph and a vertex $j$, assume one wants to investigate whether
there exists a $k$ such that ${\cal N}^k(j)=\{0,1,\ldots,N-1\}$.
Instead of a direct approach of constructing recursively the sets
${\cal N}^k(j)$ with a priori no upper bound on the number of steps,
one can use the criteria of the previous section. In particular, one
first construct an (arbitrary) quantum walk on the graph which can
be easily done and then calculates the associated reduced
connectivity graph. The connectedness of this graph is equivalent to
the existence of the previous value of $k$.

In conclusion,  there are four  main things which we have
accomplished in this section: 1) An explicit constructive control
technique. 2) An
 upper bound on the number of steps needed for any state transfer.
 3) A controllability criterion based on the sets ${\cal N}^k(j)$
 of vertices that can be reached on the graph in $k$ steps.4) An equivalence between
  this criterion and  the one in the previous section.

\section{SOME EXAMPLES}
\label{deg2}

\subsection{Graphs of degree 2}

The simplest non-trivial example are quantum walks on cycles, i.e.,
graph of degree $2$. The controllability for the fully centralized
case, i.e., with the coin operation identical for every vertex was
studied  in \cite{DGF} and generalized to lattices in \cite{FD}. Let
us denote by $|+\rangle$ and $|-\rangle$ an orthonormal basis of the
bi-dimensional coin space ${\cal{C}}$. Thus the coin tossing
operation will be of the form (\ref{definition-1}) with $Q_j\in
U(2)$, and the conditional shift will be of the type: \be{grado2-2}
S=\left(
\begin{array}{cccc}
    P_+  &  0    \\
    0   & P_-
     \end{array} \right).
 \ee
 Here $P_+$ and $P_-$ are two  matrices  representing
 the permutations associated with  the two coin
 values $+$ and $-$, respectively.
 The possible quantum walks on the cycle are described in the
 following proposition.

%%%%%%%%%%%%%%%%%%%%%%%%%%%%%%%%%%%%%%%%%%%%%%%%%%%%%%%%%%%%%%%%%%%
 %shows that,  under the connectedness  assumption
 %on the graph, when the number of nodes  $N$ is odd
 %necessarily $P_+$ is the permutation matrix representing
 %  a complete cycle and $P_-$ is its inverse. On the other hand,  while when $N$ is even
 % to the previous possibility we have to add the
 % case where $P_+$ and $P_-$ are matrices
 % representing a sequence of $N/2$ exchanges.
%%%%%%%%%%%%%%%%%%%%%%%%%%%%%%%%%%%%%%%%%%%%%%%%%%%%%%%%%%%%%%%%%%%

 \bp{grado2-3}
%Under the assumptions H1), H2), and (\ref{definition-3})
If $d=2$ then the matrices $P_+$ and $P_-$ of equations
(\ref{grado2-2}) are  necessarily of the following form:
\begin{itemize}
\item[a)] $P_+$ is the matrix representing a complete cycle, $\sigma_+$  i.e.,
(after possibly relabeling the vertices) $\sigma_+:=(012\cdots N-1)$
and $P_-=P_+^{-1}$.
\item[b)]  $P_+$ and $P_-$ are the matrices representing permutations $\sigma_+$ and $\sigma_-$, respectively,
 that are sequences of exchanges of two
 adjacent symbols,  i.e., (after possibly relabeling the vertices)
$\sigma_+:=(01)(23)\cdots (N-2 \, N-1)$, $\sigma_-:=(12)(34)\cdots
(N-3 \, N-2)(N-1 \,0)$.  This is possible only when $N$ is even.
\end{itemize}
 \ep

  \bpr

Let  $\sigma_+$ be the permutation on the nodes given by the matrix
$P_+$.
% We first show that
% \begin{quote}
% (\#) \
%  $\sigma_+$ is either a sequence of exchanges or a compete cycle.
%  \end{quote}

Write $\sigma_+$ as a sequence of cycles, $(01\cdots r_1)(r_1+1 \,
\cdots \, r_1+r_2) \cdots (r_1+r_2+\cdots r_k \cdots N-1)$, for
$k\geq 1$.  Since by assumption H2) we do not have self-loops, all
cycles must have length $\geq 2$. If all cycles are of length $2$,
then we have a sequence of $\frac{N-1}{2}$ exchanges, and we must
necessarily have that $N$ is even. Assume now that there exists a
cycle of order $p>2$, therefore, modulo a possible relabeling of the
vertices, we have
\[
\sigma_+=(01 \cdots {p})\sigma'.
\]
 We need to show  that $p=N-1$. Assume, by the way of contradiction,
that $p<N-1$. Since the permutation $\sigma_+$ corresponds to the
edges of the graph $G$, all the nodes $\{0,1,\ldots,p\}$ must have
two edges, one connecting $i$ to ${i+1}$ and the other connecting
$i$ to ${i-1}$ ($mod \, N$). If $p < N-1$, since  $G$ is regular and of degree 2,
there
cannot be any edge connecting one of the first $p$ nodes with the remaining
nodes. This contradicts the connectedness assumption on $G$,
thus the only possibility is $p=N-1$.

Now if we are in the case where
  $\sigma_+=(01 \cdots{N-1})$, then, $\sigma_+$ corresponds to motion
  along every  edge in one direction. Necessarily $\sigma_-$ will correspond
  to motion along the edges in the opposite direction, i.e., $\sigma_-=\sigma_+^{-1}$.

  On the other hand, assume that $\sigma_+$ is a sequence of
  exchanges, and let ${\sigma}_-$  be the permutation corresponding to  $P_-$. By
  repeating the same argument as before, we conclude that
 ${\sigma}_-$ is either   a sequence of exchanges or a
 complete cycle. However the last choice is not possible
 otherwise the permutation given by $\sigma_+$ would have to be
 its inverse, which is again a complete cycle. By examining the
 graph, it also follows that if $\sigma_+:=(01)(12)\cdots (N-2 \,
 N-1)$, then
$\sigma_-:=(12)(34)\cdots (N-3 \, N-2)(N-1 \,0)$.

\epr

As we have seen in Theorem \ref{independence} the controllability of
the quantum walk does not depend on the particular walk considered
but only on the graph. According to the previous proposition, in the
case $N$ odd we have only one possible type of quantum walk, while
in the case $N$ even, for the same $N$ there may be two
non-isomorphic walks. However their controllability properties
should coincide according to Theorem \ref{independence}. Let us
treat the case {\bf $N$ odd} first. Applying the criterion of
Algorithm 1 we calculate the permutation $\sigma_-^{-k}\sigma_+^{k}$
for some $k$. for $k=1$, we obtain \be{keq1}
\sigma_-^{-1}\sigma_+=\sigma_+^2=(0\,2\,4\,\cdot \, N-1
\,1\,3\,\cdots N-2),  \ee which is a full cycle. Therefore the
reduced connectivity graph is connected and the system is
controllable. Alternatively, we can apply the test of Theorem
(\ref{con-constructive}). Consider the node $0$ and the associated
sets ${\cal N}^k(0)$. We have that ${\cal
N}^{N-1}(0)=\{0,1,2,\ldots,N-1\}$. In order to see this order the nodes
of the cycle in clockwise order from $0$ to $N-1$. {}From $0$  it is
possible to reach in $N-1$ steps any node $0$, $2$, $\ldots$, $N-1$.
To see this notice that for $j=0,\ldots,\frac{N-1}{2}$, we can reach
the node $N-1-2j$ by moving $j$ times between $0$ and $1$ (so having
$2j$ steps) plus performing $N-1-2j$ additional steps clockwise.
Analogously, one can see that $\{ 1,3, \ldots , N-2\}$ are in ${\cal
N}^{N-1}(0)$. To reach $1+2j$, for $j=0,1,\ldots,\frac{N-3}{2}$ in
$N-1$ steps, one can move  $j$ times between $0$ and $1$ (and this
gives $2j$ steps) and then move counterclockwise with $N-1-2j$
additional steps. It is also easy to see that $N-1$ is the minimum
$k$ so that ${\cal N}^k(0)=\{0,1,\ldots,N-1\}$ and this minimum
value would be the same if we considered another node instead of
$0$. Therefore ${\bf k}$ in (\ref{defkappa}) is $N-1$ and since
$r=N$ in this case the upper bound on the number of steps given by
Theorem (\ref{con-constructive}) is $2(N-1)+N$. One can in fact get
a better bound since in this case the conditions described in Remark
\ref{minimizer} are verified with $$\tilde
C_1:=\left(\begin{matrix}0 & 1 \cr -1 & 0\end{matrix}\right) \otimes
I \texttt{ and } \tilde C_2=\tilde C_1^{-1}:=\left(\begin{matrix}0 &
-1 \cr 1 & 0\end{matrix} \right) \otimes I.$$

Extensions of the controllability result can be obtained.  Applying
Proposition \ref{propoprod} one has that $p$- dimensional lattices
with on odd number of vertices in every dimensions necessarily give
rise to controllable quantum walks.

\vs

For the case {\bf $N$ even}, consider first the case where the two
permutations $\sigma_+$ and $\sigma_-$ are full cycles. Applying the
criterion of Algorithm 1 we study the permutations
$\sigma_-^{-k}\sigma_+^{k}=\sigma_+^{2k}$ one sees that for every
$k$, $\sigma_+^{2k}$ is given by two cycles of length $\frac{N}{2}$
each containing only even or odd numbered vertices. Therefore the
reduced connectivity graph has two connected components each with
$\frac{N}{2}$ vertices and the system is not controllable. The
dynamical Lie algebra is the direct sum of two $su(N)$ according to
Theorem \ref{structureL}. If we apply  the criterion of Theorem
\ref{con-constructive} we find that ${\cal N}^k(0)$ contains only
even (odd) numbered nodes for $k$ even (odd) and this implies that
the system is not controllable. In the remaining case, an
application of Algorithm 1 gives the same dynamical Lie algebra and
using the criterion of Theorem (\ref{con-constructive}) gives the
same sets ${\cal N}^k$ (the criterion is independent of the walk and
the graph is the same).

\subsection{Example of a controllability algorithm}
\label{esempio-algoritmo}

Consider the quantum walk whose graph is given in Figure 1. The
graph has $6$ nodes and degree $d=3$, thus any associate quantum
walk has state space of dimension $18=6\cdot 3$.

\begin{figure}
\begin{center}
$ \xymatrix{ & & &  {{\bf{0}}} \ar@{-}@/_1pc/[dll]_<{- \ \ } _>{\ \
+  } \ar@{-}@/^1pc/[drr]^<{\  \ +}^>{-\ \ }
 \ar @{-}@/_1pc/[ddd]_<{\ \ c } ^>{\ c  } & & &  \\
 &  {{\bf{5}}} \ar@{-}@/_/[d]  _<{\ \ -}_>{\ \ + }
 \ar@{-}@/_/[rrrr]  ^<{\ \ c}^>{ c \ \  }& & &  &   {{\bf{1}}} \ar@{-}@/^/[d] ^<{+ \ \ }^>{  - }  &\\
  &  {{\bf{4}}} \ar@{-}@/_/[drr]  _<{\ \ -}_>{ + \  \ }
   \ar@{-}@/_/[rrrr]  ^<{\ \ c}^>{ c \ \  }& & &  &   {{\bf{2}}} \ar@{-}@/^/[dll] ^<{+ \ \ }^>{\ \ -}& \\
  & & &  {{\bf{3}}}  & & &  \\
} $ \caption{Graph with $N=6$   and $d=3$  }
\end{center}
\end{figure}
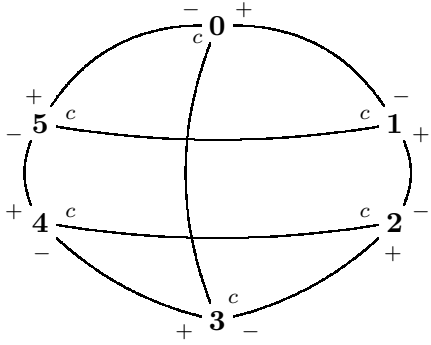

For this graph, it is easy to see that we have:
\[
\begin{array}{lcl}
{\cal N}^1(0) &= & \{ 1,\ 3,\ 5 \}, \\
{\cal N}^2(0) &= & \{ 0, \ 1,\ 2,\ 4, \ 5 \}, \\
{\cal N}^3(0) &= &\{ 0, \ 1,\ 2,\ 3, \ 4, \ 5 \}.
\end{array}
\]
This fact implies that
  any quantum walk on this graph will be completely controllable.
  Let us consider the problem to steer the initial state
\begin{equation}\label{initstate}
  |\psi_0\rangle= |+\rangle\otimes |0\rangle,
  \end{equation}
   i.e.,  a state where the probability is concentrated in the $0$ node,   to a final state
$|\psi_f\rangle$ with the probability uniformly  distributed among
all the nodes, i.e., $|\psi_f \rangle$ of the form
\begin{equation}\label{desstate}|\psi_f\rangle =\frac{1}{\sqrt{6}}
\sum_{j=0}^5 |c_j\rangle \otimes |j \rangle \end{equation}
 where
$|c_j \rangle$ are general (not necessarily basis) states in ${\cal
C}$.

We assume, as described in the picture,  that  the two coin values
$|+\rangle$ and $|-\rangle$ correspond to permutations
$P_{+}=(012345)$ and $P_{-}=(054321)$ while with  the third coin
value, which will be denoted by $|c\rangle$, we associate the
permutation $P_{c}=(03)(15)(24)$. We proceed  by using  the procedure described in
Proposition \ref{prima-nodi}.  First consider ${\cal N}^3(0)$.
\[
{\cal N}^3(0) =\{ 0, \ 1,\ 4,\ 2, \ 3, \ 5 \}=
\]
\[ = \{ P_+(5), P_c(5),
P_-(5), P_c(4), P_-(4), P_+(4)\}.\footnote{Thus, using the notations of Proposition \ref{prima-nodi}, here we have $z_1=5$ and $z_2=4$.
Notice that this choice is not unique, in fact for example $1=P_c(5)=P_+(0)$.
Any possible choice will lead to   different sequence of coin tossing operations.}
\]
The expression suggests that if we were in a state
\begin{equation}\label{almost}
|\psi_2\rangle=\frac{1}{\sqrt{2}} |c_4 \rangle \otimes |4 \rangle +
\frac{1}{\sqrt{2}} |c_5 \rangle \otimes |5 \rangle,
\end{equation}
and applied a coin operation \begin{equation} \label{coinop} Q_5
\otimes |5 \rangle \langle 5|+Q_4 \otimes |4 \rangle \langle 4|+
{\bf 1}_3 \otimes ({\bf 1}_6-|5 \rangle \langle 5|-|4 \rangle
\langle 4|),
\end{equation}
with $Q_5$ ($Q_4$) a unitary transformation mapping $(c_5\rangle$
($|c_4 \rangle$) to $\frac{1}{\sqrt{3}}(|+\rangle
+|-\rangle+|c\rangle)$ we would obtain  state of the form
(\ref{desstate}). Therefore the problem is reduced to obtain a state
of the form $|\psi_2\rangle$ in (\ref{almost}). To do that we
examine $4$ and $5$ in ${\cal N}^2(0)$ and we have $4=P_-(5)$ and
$5=P_c(1)$. This suggests that if we have a state
\begin{equation}\label{olp} |\psi_1\rangle:= \frac{1}{\sqrt{2}}
|d_5\rangle \otimes |5 \rangle +\frac{1}{\sqrt{2}} |d_1 \rangle
\otimes |1 \rangle,
\end{equation}
we could transfer to a state of the form (\ref{almost}) by applying
a coin transformation depending on the walker which maps
$|d_5\rangle $ into $ |+\rangle$ and $|d_1\rangle $ into $|c\rangle$
followed by a conditional shift. Finally, examining $5$ and $1$
which are in ${\cal N}^1(0)$, we have that $5=P_-(0)$ and
$1=P_+(0)$. Starting from a state $\psi_0$ in (\ref{initstate}) and
applying a coin transformation mapping $|+\rangle$ into
$\frac{1}{\sqrt{2}} |- \rangle + \frac{1}{\sqrt{2}} |+\rangle$
followed by a conditional shift $S$, we obtain the state in
(\ref{olp}). The procedure to go from $|\psi_0\rangle$ to
$|\psi_f\rangle$ applies the above procedure in reverse.

%%Say that Control Algorithms can be seen as information theory algorithms
%%

%\subsection{Future Works}

%%%%%%%%%%%%%%%%%%%%%%%%%%%%%%%%%%%%%%%%%%%%%%%%%%%%%%%%%%%%%%%%%%%%%%%%%%%%%%%%
\section{ACKNOWLEDGMENTS}

D. D'Alessandro research was supported by NSF under Grant No.
ECCS0824085. D. D'Alessandro also acknowledges  the kind hospitality
by the Institute for Mathematics and its Applications (IMA) in
Minneapolis where  this work was performed.

%%%%%%%%%%%%%%%%%%%%%%%%%%%%%%%%%%%%%%%%%%%%%%%%%%%%%%%%%%%%%%%%%%%%%%%%%%%%%%%%

\section*{Appendix A:Further remarks on the structure of
the dynamical Lie algebra ${\cal L}$. }
 In this short appendix, we give a graph theoretic argument to show  that
 the the number $m$ of connected components of the reduced
 controllability graph in Theorem \ref{structureL} can only be 1 or
 2. In order to see this,  define an equivalence relation $\thicksim$ on the
set of vertices $V$ saying that $a \thicksim b$ if there exists a
path of even length connecting $a$ and $b$. The partition of the set
$V$ considered in Theorem \ref{structureL} corresponds to partition
in equivalence classes with respect to this equivalence relation
according to the discussion preceding Theorem \ref{equivale}. Now,
fix a $j\in V$ and consider a set $V_o (j)$ as the set of vertices
that can be reached by $j$ in an odd number of steps and a set $V_e
(j)$ of vertices that can be reached in an even number of steps.
Clearly $V=V_o(j) \bigcup V_e(j)$. Moreover if $a$ and $b$ are in
$V_o(j)$ (or $V_e(j)$), $a \thicksim b$. Therefore either $V=V_o(j)=
V_e(j)$ or $V_o(j)$ and $V_e(j)$ ar disjoint and they give two
connected components in the reduced connectivity graph. This
discussion shows that the example of the cycle discussed in Section
\ref{deg2} is somehow prototypical. It also shows that another
equivalent condition of controllability is that given a $j \in V$ we
are able to find a vertex which we can reach in both an odd and an
even number of steps.


\begin{thebibliography}{99}


\bibitem{Notcontr} F. Albertini and D. D'Alessandro, Notions of
controllability for bilinear multilevel quantum systems, {\it IEEE
Transactions on Automatic Control}, 48, No. 8, 1399-1403 (2003).

\bibitem{FD}
F.  Albertini and D. D'Alessandro,  "Analysis  of Quantum Walks with Time-Varying Coin on
$d$-Dimensional Lattices", {\it Journal of Mathematical Physics,}
50, 122106 (2009).

\bibitem{Ambainis} A. Ambainis, Quantum walks and their algorithmic
applications,  {\it International. Journal of Quantum Information},
{\bf 1}, 507?518, (2003).


\bibitem{Ambaal} A. Ambainis, J. Kempe, and A. Rivosh, Coins make
quantum walks faster, Proc. 16th ACM-SIAM SODA, p. 1099-1108 (2005).


\bibitem{Childs} A. M. Childs, On the relationship between
continuous and discrete-time quantum walk, {\it Communications in
Mathematical Physics} 294, 581?603 (2010)


\bibitem{Mikobook} D. D'Alessandro, {\it Introduction to Quantum
Control and Dynamics}, CRC-Press, Boca Raton FL, 2007.

\bibitem{MikoConDis}D. D'Alessandro, Connection Between Continuous and Discrete Time
Quantum Walks; {}From $d$-Dimensional Lattices to General Graphs, to
appear in {\it Report on Mathematical Physics}.

\bibitem{Mikogenmeth}D.  D'Alessandro , General methods to control right-invariant
systems on compact Lie groups and multilevel quantum systems,   2009
J. Phys. A: Math. Theor. 42 395301

\bibitem{DGF}
D. D'Alessandro, G. Parlangeli and F.  Albertini, "Nonstationary
quantum walks on the cycle", {\it J. Phys. A: Math. and Theor.}
(2007), 40, 14447-14455.


\bibitem{Helgason} S. Helgason, {\it Differential geometry, Lie groups
and symmetric spaces}, Academic Press, New York,  1978.


\bibitem{MHillery} M. Hillery, D. Reitzner and V. Bu$\breve{z}$ek,
Searching via walking: How to find a marked subgraph of a graph
using quantum walks, xxx.arXiv:quant-ph 0911.1102v1

\bibitem{Kempe} J. Kempe, Quantum random walks - an introductory
overview, {\it Contemporary Physics,} Vol. 44 (4), p. 307-327, 2003.

\bibitem{Kendon} V. Kendon, Decoherence in quantum walks - a review,
{\it Math. Struct. in Comp. Sci} 17(6) pp 1169-1220 (2006)

%\bibitem{Krovi} H. Krovi and T. A. Brun, Quantum walks on quotient
%graphs, {\it Physical Review A}, 75, 062332, (2007).

\bibitem{MITguy} M. Mohseni, P. Rebentrost, S. Lloyd and A.
Aspuru-Guzik, Environment-assisted quantum walks in photosynthetic
energy transfer, {\it Journal of Chemical Physics} 129, 174106
(2008).


\bibitem{SagleWalde} A. A. Sagle and R. E. Walde, {\it Introduction to
Lie Groups and Lie Algebras},  Academic Press, New York,  1973.

\bibitem{Sepanski} M.R. Sepanski, {Compact Lie Groups}, Graduate
Texts in Mathematics, Vol. 235, Springer 2007.

\bibitem{Strauch}  F.W. Strauch, Connecting the discrete and the
 continuous-time quantum walks, Phys. Rev. A 74, 030301 (R) (2006).


\bibitem{Tulsi} A. Tulsi, Faster quantum walk algorithm for two
dimensional spatial search, {\it Physical Review A}, 78, 012310
(2008).


\bibitem{TuriniciRabitz} G. Turinici and H. Rabitz, Wavefunction
controllability for finite-dimensional bilinear quantum systems,
{\it J. Phys. A: Math. Gen.}, { 36},  2565-2576, (2003).


\end{thebibliography}
\end{document}